\documentclass[%
reprint,
amsmath,amssymb,
prb,
]{revtex4-2}
\usepackage[title]{appendix}
\usepackage{graphicx}
\usepackage{dcolumn}
\usepackage{bm}
\usepackage{gensymb}
\usepackage{float}
\usepackage{multirow}
\usepackage{tabularx}
\usepackage{amsmath}
\usepackage{color,soul}
\usepackage{makecell}
\usepackage{subcaption}
\usepackage{diagbox}
\usepackage{physics}
\usepackage[colorlinks,allcolors=black,pdftex]{hyperref} 
\usepackage[capitalise]{cleveref}
\usepackage[dvipsnames]{xcolor}

\newcommand{\rcol} {\textcolor{red}}

\usepackage{graphicx} 
\usepackage{hyperref}

\begin{document}

\preprint{APS/123-QED}


\title{Light induced Berezinskii-Kosterlitz-Thouless transition in Superconducting Films}

\author{Tien-Tien Yeh$^{1}$, Evan Wilson$^{1}$, Mikael Fogelström$^{2,3}$, 
Alexander Balatsky$^{1,2}$}
\affiliation{$^{1}$Department of Physics, University of Connecticut, Storrs, Connecticut 06269, USA}
\affiliation{$^{2}$Nordita, Stockholm University, and KTH Royal Institute of Technology, Hannes Alfvéns väg 12, SE-106 91 Stockholm, Sweden}
\affiliation{$^{3}$Department of Microtechnology and Nanoscience MC2, Chalmers University of Technology, SE-41296 Göteborg, Sweden}

\begin{abstract}

We report a light-driven non-equilibrium vortex Berezinskii-Kosterlitz-Thouless (BKT) transition in a superconductor. We use a time-dependent Ginzburg-Landau model to demonstrate vortex–antivortex deconfinement via light induced fields. The transformation occurs independently of thermal fluctuations and is viewed as a quantum phase transition. The resulting phase map mirrors QCD phase diagram, delineating confined, premelted, and fully deconfined vortex phases. The nature of these phases is discussed. Transitions between phases are governed by light induced depairing and phase fluctuations, establishing a new class of light-induced topological transitions.

\end{abstract}

\maketitle


Topological excitations, such as vortices in superconductors or XY magnets, are absent in the ordered state that favors the smooth superconducting phase configurations. Hence, the trivial ground state does not exhibit visible topological excitations~\cite{kibble1976topology,mermin1979topological}. Following the pioneering work of BKT,  we know that smooth configurations often hide confined topological excitations such as soliton-antisoliton pairs, domain walls, and vortex-antivortex pairs~\cite{kosterlitz1973ordering,berezinskii1971destruction,nelson1979dislocation}. Given the right conditions, the confinement-deconfinement transition can be induced and topological excitations are liberated at higher temperatures or higher densities of matter. The list of topological deconfinement transitions now includes  examples of superconductors, magnets~\cite{castelnovo2008magnetic,shuryak1982role,rajaraman1979solitons,kosterlitz1973ordering,nelson1979dislocation,coleman1975quantum}, and cold atoms~\cite{cheng2022tunable,tagliacozzo2013simulation,banerjee2012atomic}. We also mention a related example of confinement-deconfienment transition in quantum chromodynamic matter (QCD), a quark-gluon plasma, ~\cite{gupta2011scale,fischer2009deconfinement,aoki2006qcd,stephanov2004qcd}.  
To date, majority of the investigated cases concern equilibrium phases ~\cite{ranabhat2025dynamical,kosterlitz1973ordering,berezinskii1971destruction,nelson1979dislocation}. 
We are now in a position to probe similar transitions out of equilibrium  where confinement-deconfinement phases are   induced dynamically using drives such as light~\cite{dagvadorj2015nonequilibrium}.

\begin{figure*}[!htbp]
    \centering
    \includegraphics[width=0.9\textwidth]{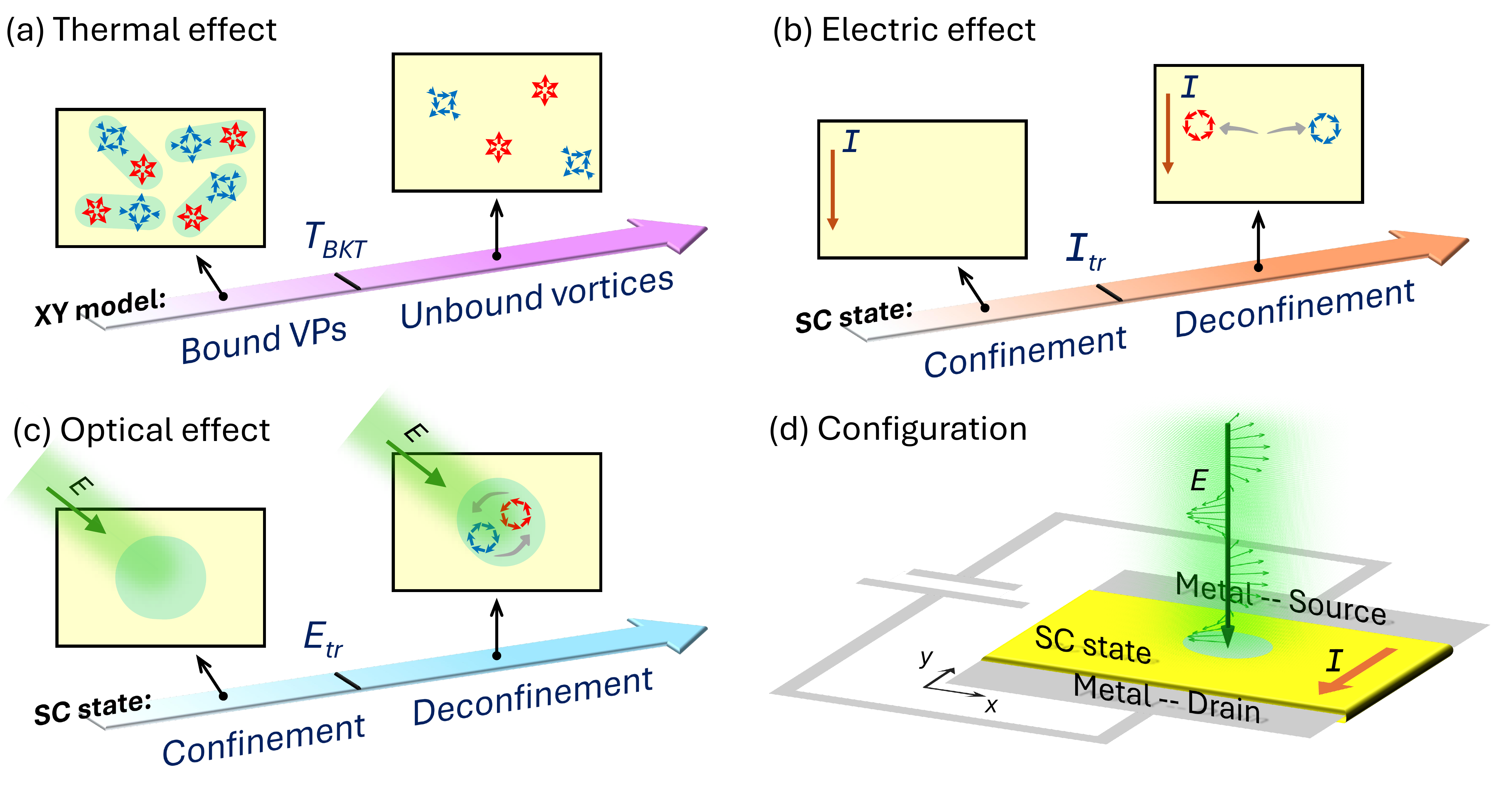}
    \caption{Schematic diagrams illustrating (a) the BKT phase transition of vortices in the XY model, (b) confinement–deconfinement of vortices in a superconducting (SC) state induced by electric current, and (c) the one induced by light. These three types of phase transitions occur at distinct critical values of control parameters: the transition temperature $T_{BKT}$, the transition current $I_{tr}$, and the transition optical field $E_{tr}$, respectively. Yellow regions represent the unperturbed SC order (see the color bar for suppression of SC order. Red and blue arrows   regions denote vortices and antivortices, respectively. (d) Simulation configuration: the superconducting slab is placed in the $xy$-plane, with a circularly polarized light beam incident normally at its center. To introduce the DC current, the upper and lower edges of the slab are connected to a metal source and metal drain, respectively.}
    \label{fig:config}
\end{figure*}

The goal of this letter is to uncover the phase transition that governs the deconfinement of vortices at $T=0$ in a light-driven superconductor (SC). We propose a non-equilibrium transition mechanism in which optically generated vortex–antivortex pairs (VP) are propelled by a direct-current (DC) bias. We show that SC can undergo a dynamically induced topological phase transition from a confined state to a deconfined state. As illustrated in~\cref{fig:config}, this transition resembles vortex unbinding in the XY model, but originates from a fundamentally different process; where the stirring of the dynamical phase of a charged SC is produced by light. Building on our earlier studies on quantum printing~\cite{LG_TDGL_I,LG_TDGL_II}, which examined optically nucleated SC VPs revolving around the laser spot, we now show that the optical driving can induce a deconfined transition even at nominal $T=0$ with deconfined states, A new intermediate phase the so called {\it premelted phase} that enables VP unbinding within the illumination region; and thus establish a natural extension of the BKT transition toward a QCD-like 2D phase diagram of quantum phases with no regard for temperature.

\begin{figure*}[!htbp]
    \centering
    \includegraphics[width=0.9\textwidth]{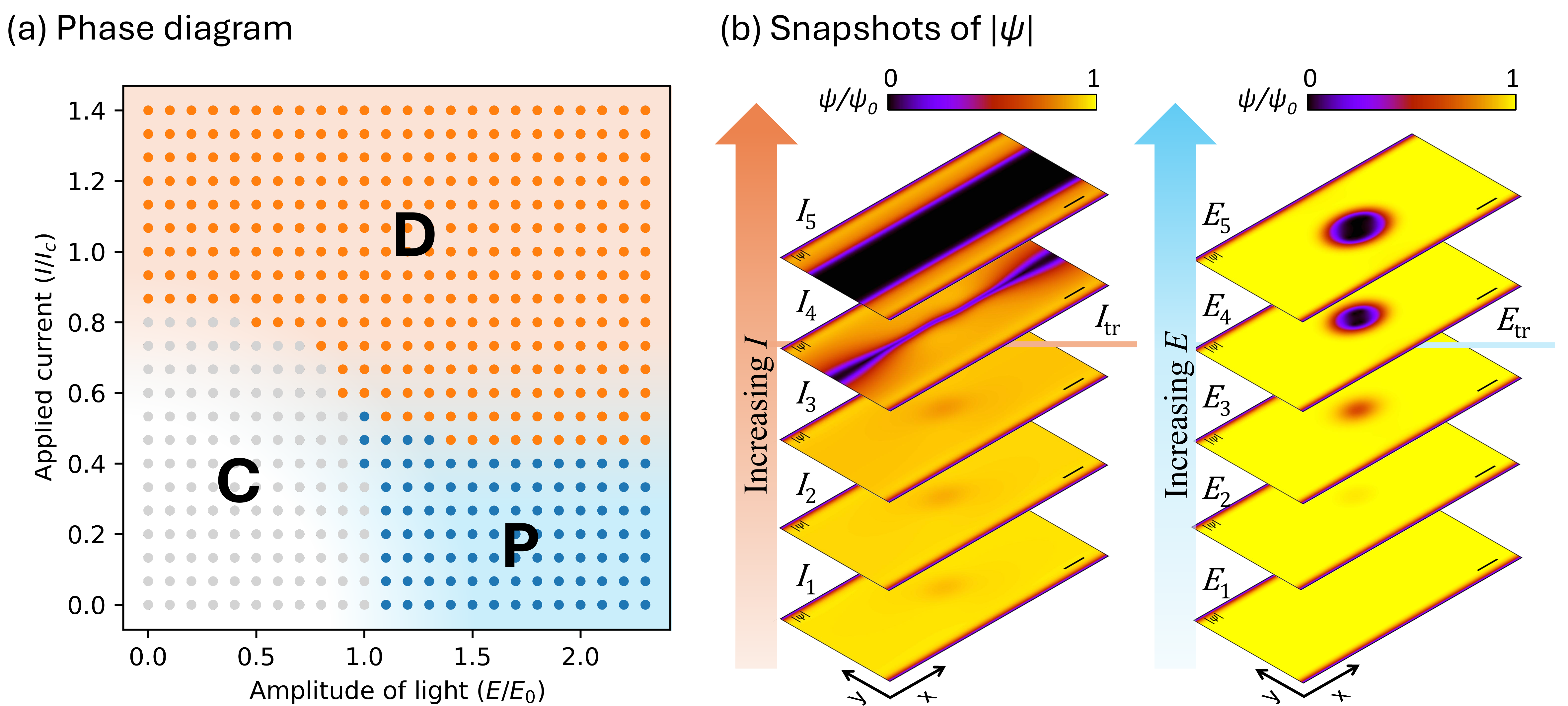}
    \caption{
    (a) $I$–$E$ phase diagram of confinement-deconfinement transitions in vortices. The gray, orange, and blue data points, marked ``$\bf{C}$'', ``$\bf{D}$'', ``$\bf{P}$'' represent the confinement, deconfinement, and premelted phases, respectively. Complete results for the dynamics in respective phase are provided in the Supplementary Materials~\cite{supp_PD}.
    (b) Snapshots of $\abs{\psi}$ profiles at $t = 500 \tau_{GL}$ for different applied currents and amplitudes of light field. For the current-dependent series (fixed $E = 0.5E_0$), the applied currents are $I_1 = 0.59I_c$, $I_2 = 0.66I_c$, $I_3 = 0.73I_c$, $I_4 = 0.80I_c$, and $I_5 = 0.86I_c$. For the light-dependent series (no applied current), the amplitudes are $E_1 = 0.0E_0$, $E_2 = 0.5E_0$, $E_3 = 1.0E_0$, $E_4 = 1.5E_0$, and $E_5 = 2.0E_0$.
    }
    \label{fig:PD}
\end{figure*}


To investigate the simultaneous effects of laser beam and applied DC current on SC order parameter and vortex dynamics, we employ the generalized time-dependent Ginzburg–Landau (gTDGL) model~\cite{kramer1978theory, watts1981nonequilibrium}.
The dimensionless gTDGL equation can be written as
~\cite{LG_TDGL_I, bishop2023pytdgl,berdiyorov2009kinematic,kramer1978theory, watts1981nonequilibrium},
\begin{multline} 
\label{eq:TDGL}
\frac{u}{\sqrt{1+\gamma^2\left| \psi \right|^2}}
\left(\frac{\partial }{\partial t}+i\mu+\frac{\gamma^2}{2}\frac{\partial \left| \psi \right|^2}{\partial t}\right)\psi
\\ =(1-\left| \psi \right|^2)\psi+(\nabla -i\boldsymbol{A})^2\psi ,
\end{multline}
where $\psi$ is the complex order parameter, i.e., $\psi=|\psi|\exp{i \theta_{s}}$ with phase field $\theta_s$, the parameters $\mu$, $\boldsymbol{A}$, $u$, $\gamma$ represents scalar potential, vector potential $\boldsymbol{A}$, the constant relative to the relaxation time ratio of the supercurrent and $\psi$, and the parameter associated with inelastic scattering processes which $\gamma=10$, respectively. 
In this simulation, the total current density is given by
$\boldsymbol{J}=\boldsymbol{J_{s}}+\boldsymbol{J_{n}}=\text{Im}[\psi^{*} (\nabla -i\boldsymbol{A})\psi] - \nabla\mu - \frac{\partial \boldsymbol{A}}{\partial t}$,
where $J_s$ and $J_n$ denote the supercurrent density and normal current density, respectively. 

The configuration is illustrated in ~\cref{fig:config}(d). It consists of a superconducting slab with dimensions $60\xi_0 \times 40\xi_0 \times 2\xi_0$, where $\xi_0$ denotes the coherence length. The laser beam is circular polarization, Gaussian distribution, and normally incidence at the center ($x=0, y=0$) of the slab, i.e., $\mathbf{E}(\mathbf{r},t)=E\cdot  e^{-\mathbf{r}^2/2w_0^2} \cdot e^{-i\omega t}(\hat{\mathbf{x}}+i\hat{\mathbf{y}})$, with a spot size $2w_0=15\xi_0$, frequency $40\omega_{GL}$ where $\omega_{GL}=2\pi/\tau_{GL}$ and $\tau_{GL}$ are the units of time (~\cref{app:unit}), and electric field amplitude $E$ with unit $E_0=A_0\omega_{GL}$ and $A_0$ is the unit of vector potential (~\cref{app:unit}). Units are dependent on the intrinsic properties of the material. If we consider Nb thin film at 0.99$T_c$ as an example, which is with $\xi_0=100$, then the $I_0$ and $E_0$ are expected to be $3.1$ mA and $3.8$ kV/cm, respectively. The DC normal current is applied along the $y$-axis from $+y$ to $-y$ with amplitude $I$. We utilize critical current $I_c$ as the reference for the current unit (~\cref{app:Ic}). At $x = \pm 30,\xi_0$, we impose a superconductor–vacuum boundary condition, while at $y = \pm 20,\xi_0$, we apply a superconductor–metal interface to guide the bias normal current. More details on units and boundary conditions are provided in~\cref{app:unit}.


\begin{figure}[!htbp]
    \centering
    \includegraphics[width=0.45\textwidth]{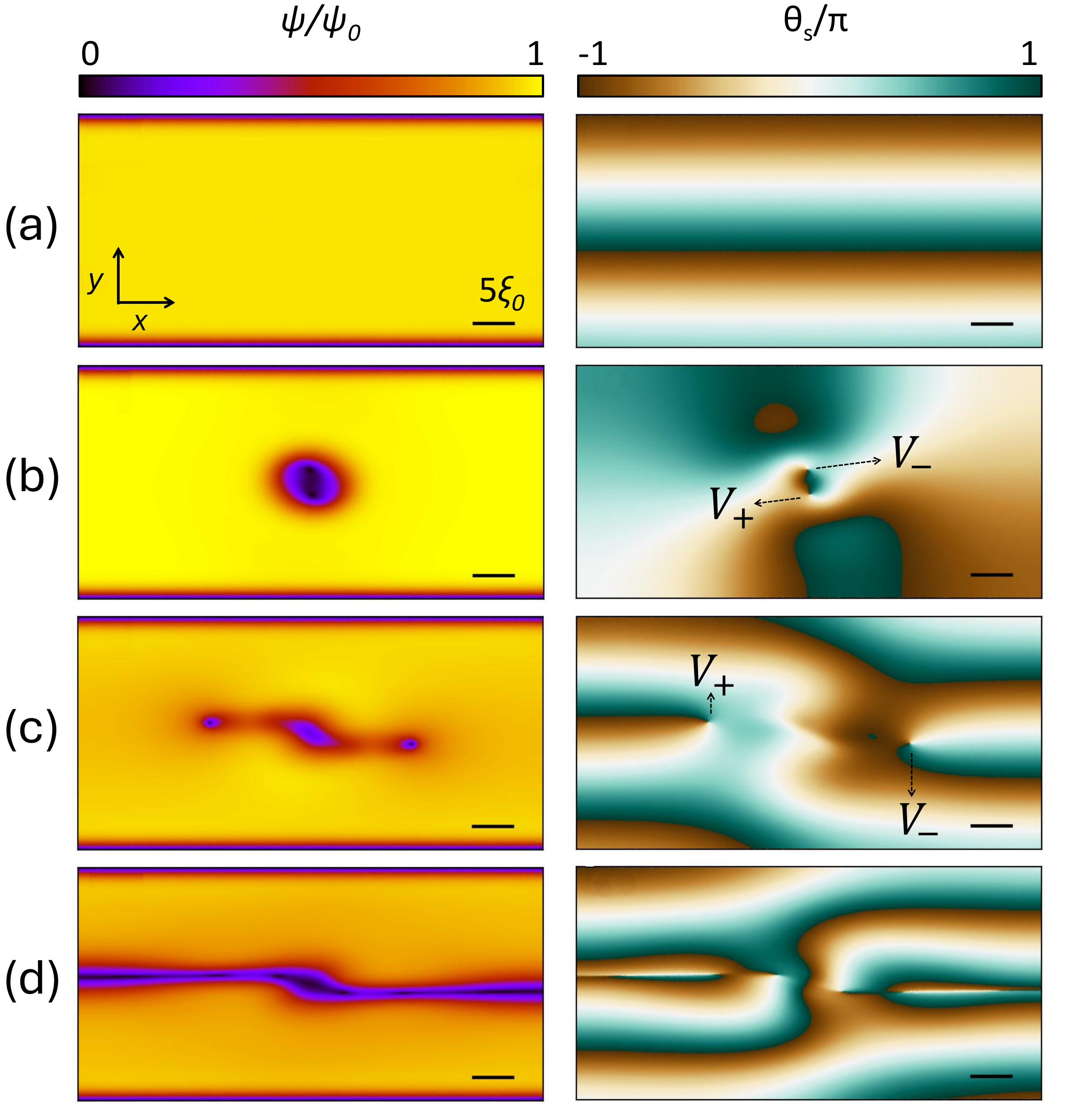}
    \caption{
    Typical profiles of the order parameter and phase field for different dynamical phases:
    (a) confined (C), (b) premelted (P), (c) deconfined (D) with VP, and (d) deconfined (D) with phase slip. These snapshots are taken from simulations with parameters $t=260\tau_{GL}$ $[I/I_c, E/E_0] =$ [0.53,0.0], [0.00,1.3], [0.60,1.1], [0.73,1.0], respectively. The labels $V_+$, $V_-$ indicate vortices, antivortices, respectively.
    }
    \label{fig:SS}
\end{figure}


To systematically investigate these phenomena, we construct an $I$–$E$ phase diagram as depicted in ~\cref{fig:PD}(a). Analogous to confinement–deconfinement diagrams observed in quark–gluon plasma studies~\cite{gupta2011scale,stephanov2004qcd}, our vortex confinement phase forms a dome-like structure originating from minimal field parameters ($E=0$ and $I=0$). 

In the simulation results, three distinct dynamical regimes emerge, governed predominantly by the optical field, DC current, or weak external fields. In the optically dominated regime, the illuminated region is premelted, leading to the depairing of SC vortices, and the formation of VPs once the optical amplitude exceeds the transition amplitude $E_{tr} \sim 1.1 E_0$ (~\cref{fig:PD}(b)). Interestingly, vortices remain confined within the optical spot, irrespective optical intensity. In contrast, under sufficiently strong DC current conditions $I_{tr}$ (dependent on $E$), vortices dynamically escape from the illuminated region, defining the deconfinement phase. If neither the optical field nor the DC current is sufficiently strong, superconducting states remain largely intact. Accordingly, we designate these three regimes as premelted (P), deconfined (D), and confined (C) phases, respectively. Notably, both premelted and deconfined phases exhibit highly dynamic, non-equilibrium vortex behavior.

The boundary between phases reveals that increasing the amplitude of the optical field $E$ progressively lowers the  current required for vortex deconfinement across the confinement–deconfinement transition. This is because the presence of the optical field assists in the depairing of the superconducting state, thereby reducing the energy required for vortex generation~\cite{LG_TDGL_I}. Below this current threshold, vortices remain constrained within the premelted region.
Similarly, the confinement–premelted phase boundary is affected by current-induced depairing effects~\cite{de2018superconductivity,bardeen1962critical}.  Growing  applied current facilitates the premelting process and the optical field strength $E$ required to induce the transition reduces further. The premelted–deconfinement phase boundary is  influenced not only by optical-field-induced depairing but also by a topological constraint related to energy costs of vortex antivortex strings, which will be discussed later. These transition boundaries reveal the interplay between optical and electrical parameters in controlling creation and dynamcis of topological excitations.
    
\begin{figure*}[!htbp]
    \centering
    \includegraphics[width=0.9\textwidth]{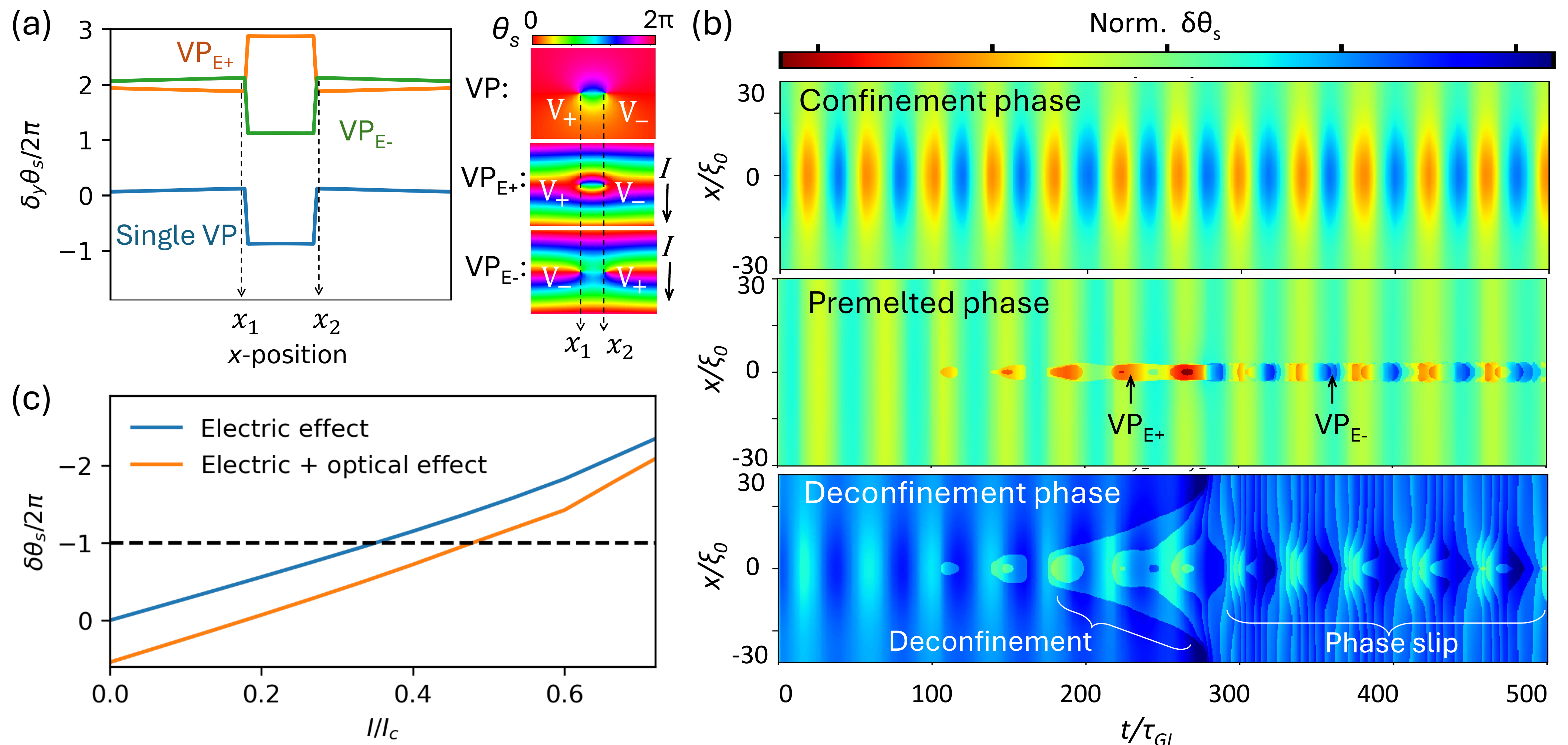}
    \caption{
    (a) Results of $\delta_y \theta_s = \theta_s(x, y_2) - \theta_s(x, y_1)$ (where $y_1$ and $y_2$ locate at the bottom and top edges of the sample, respectively), scanned along the $x$-axis. The corresponding phase fields are shown in the right insets. 
    The kinetic energies of VP$_{E+}$ and VP$_{E-}$ are 28.4 and 23.4 times greater than that of a single VP, respectively, based on Eq.~\cref{eq:Ek_approx}. The dashed lines indicate the vortex positions at $x_1$ and $x_2$, as marked in the insets.
    Insets: Idealized phase fields for a single VP, VP$_{E+}$, and VP$_{E-}$. The solid arrows indicate the direction of the applied DC current, which induces phase stripes outside the VP region corresponding to $\delta_y \theta_s = 4\pi$.
    (b) Time evolution of $\delta_y \theta_s$ from simulation datas for three distinct phases: confinement (top), premelted (middle), and deconfinement (bottom), corresponding to parameter sets [$I/I_c$, $E/E_0$] = [0.00, 0.5], [0.34, 0.5], and [0.66, 1.0], respectively. In the premelted phase, the system exhibits alternating appearances of VP$_{E+}$ and VP$_{E-}$ states over time, manifested as the red and blue bumps as labeled, respectively. 
    (c) Effective $\delta_y \theta_s$ as a function of applied current, comparing the electric effect (blue line) and the combined electric + optical effect (orange line). The dashed line at $\delta_y \theta_s = 2\pi$ indicates the topological restriction for vortex deconfinement. Accordingly, the estimated transition current $I_{tr}$ is approximately $0.35 I_c$ for the electric effect alone, and increases to $0.48 I_c$ when the optical effect is included (~\cref{app:Itr}).
    }
    \label{fig:PF}
\end{figure*}

Apart from the phase diagram, we observe distinct features in the phase field topology that reflect underlying topological constraints of transition boundary in phase diagram. As shown in~\cref{fig:SS}, the main differences between confinement, premelted, deconfinement phases become apparent. Dynamic evolution of the phases is provided in the Supplementary Materials~\cite{supp_PD}. In general, the optical drive induces temporally periodic phase field motion, and the DC drive generates spatially periodic striping. These two driving sources independently lead to strong temporal and spatial modulations in the electric field, which are clearly manifested in the correlation functions plotted in Fig. S2 (a–d). However vortices suppress these modulations, as seen in Fig. S2 (e,f).

In the phase field snapshots presented in~\cref{fig:SS}, only the confinement phase which exhibits trivial topology by the absence of vortices, the other phases present various features of singular points in the phase field. In particular, deconfined vortices and those in the premelted region exhibit markedly different phase field features. During the deconfinement process, the phase field typically displays two semi-open stripes that extend toward the sample edge, in contrast to the closed phase contours observed in the premelted region at low or zero $I$.
These distinct profiles give rise to different phase gradients, which in turn lead to different kinetic energy not only near the vortices but throughout the entire system (~\cref{app:VPEk}).

The phase field of VPs introduces a $2\pi$ phase jump between the vortex and antivortex i.e. a string, while maintaining an almost constant value outward (see ~\cref{fig:PF}(a) and ~\cref{app:theta_s}). Under an applied DC current, the orientation of the vortex pairs becomes crucial. 
This $2\pi$ phase string along the vortex pair can superpose with or suppress the locally perpendicular current, effectively resulting in the additional $\Delta\theta_s = \pm 2\pi$ between vortex and antivortex, which can increase or decrease the local kinetic energy.  This difference in energy is responsible for the Magnus force pulling vortices apart by current.  We denote these VP$_{E+}$ and VP$_{E-}$, respectively. 

In the deconfinement regime, the phase fields resemble VP$_{E-}$ in~\cref{fig:PF}(a)(b), because it released more energy during the separation of VP. These observations of vortex motion are consistent with the expected traveling direction of transverse vortices driven by the Lorentz force~\cite{bezuglyj2022vortex,misko2007negative}, and are also used to support the estimation of transition boundary between the premelted and deconfinement phases via underlying topological mechanism. Where the applied current must be sufficient to induce a total phase difference $\delta \theta_s > 2\pi$; otherwise, the supercurrent has to flow in the opposite direction in the VP region,
violating the current flow from the boundary and against the occurrence of phase slip after vortex reflection (see ~\cref{app:reflec_V}). Accordingly, the estimated $I_{tr}$ almost consists of $I_{tr}$ observed from the phase diagram (see the estimation in~\cref{fig:PF}(c) and details in~\cref{app:Itr}).

Moreover, as $I$ increases in the premelted phase, the lifetime of VP$_{E-}$ increases while that of VP$_{E+}$ decreases, revealing a gradual crossover characteristic of a dynamical second-order phase transition in the $\theta_s$ profile. This behavior is in stark contrast to the clear transition between the premelted and deconfinement phases because of topological constraint of VP.

Notably, the investigation of vortex deconfinement offers a novel, non-thermal and non-magnetic route for generating, preserving, and manipulating SC vortices. When deconfinement occurs without inducing a phase slip, controlled by $I$, the resulting vortices can exhibit  extended lifetimes.
To explore this, we simulate a scenario in which the light is switched off after VP generated, under the condition $[I/I_c, E/E_0] = [0.53, 1.5]$. The simulation results show that the vortices persist for hundreds of times longer than the vortex-antivortex recombination time, even exceeding the duration of the simulation. This suggests the emergence of a long-lived metastable state ~\rcol{~\cite{supp_step}}.
These findings illustrate a mechanism where the optical field serves as a trigger for vortex generation, while a weak, moderate DC current ($I < I_{tr}$ at $E = 0$) sustains and transports the vortices even after the light is removed. 

In summary, we have demonstrated a novel non-equilibrium topological phase transition in SC films, where vortex-antivortex pairs undergo confinement-deconfinement transitions induced by the interplay of optical and electrical driving. Using time-dependent Ginzburg-Landau simulations, we constructed a comprehensive $I$-$E$ phase diagram revealing three distinct regimes of confined (C), premelted (P), and deconfined (D) phases. Each region is characterized by different phase-field dynamics and topological responses. We thus uncovered a new type of non-thermal, non-magnetic BKT-like transition. This topological transition is controlled by the total phase gradient (global current)  and light fluence.  It is controlled by energy  threshold set by the interplay of current-driven phase strings and optically induced premelting.
Our results introduce a fundamentally different mechanism for topological phase control in superconductors, enabling vortex generation. The ability to separate vortex creation (by light) from vortex manipulation (by subcritical current) offers a pathway for low-dissipation control of SC states. Proposed light assisted BKT transition mechanism paves the way for new device concepts, including vortex-assisted quasiparticle propagation, fluxonic logic~\cite{wustmann2020reversible}, vortex channel for non-volatile superconducting memory~\cite{pot2023nonvolatile} and controlled quasiparticle transport~\cite{bulaevskii2011vortex} and vortex electronics~\cite{keren2023chip,golod2015single,kalashnikov2024demonstration}, 

These results may support future advances in vortex-assisted  SC topological solitons. Moreover, we anticipate that through the flexible control of current paths and sample geometry,  one might be able to steer vortices   along desired routes, realizing a vortex-conducting channel similar to an optical fiber.

\section{Acknowledgment}
 We are grateful to SZ. Lin, H. Yerzhakov
, J. Heath, P. Wong, Z. White, Y. Liu, D. Kang for useful discussions. This work was supported by the US DOE DE-SC0025880. MF is supported by Nordita. 

\appendix
\begin{appendices}


\section{\label{app:unit}Simulation details.}

In this work, we adopt the dimensionless gTDGL equation using the Python-based pyTDGL code developed by L. Bishop-Van Horn~\cite{bishop2023pytdgl}, incorporating structured light as implemented in our previous works~\cite{LG_TDGL_I,LG_TDGL_II}. The optical parameters are chosen following the settings used in Refs.~\cite{LG_TDGL_I,LG_TDGL_II}.

The length and time units of this dimensionless gTDGL model are determined by the superconducting coherence length $\xi_0$ and the supercurrent relaxation time $\tau_{GL}$ ~\cite{LG_TDGL_I,bishop2023pytdgl,berdiyorov2014dynamics,kopnin2001theory}, respectively. The unit of current $I_0$ is given by $V_{\mathrm{SC}} J_0$, where $V_{\mathrm{SC}}$ is the volume of the superconducting slab and $J_0$ is the unit of current density. Refs.~\cite{LG_TDGL_I,bishop2023pytdgl,kopnin2001theory} provide the complete formula and detailed discussion of the unit. 

For the superconductor–vacuum interface, it is implemented via Neumann boundary conditions for both $\psi$ and $\mu$~\cite{bishop2023pytdgl,jonsson2022current}, i.e. $\bf{\hat{n}} \cdot (\nabla-\it{i}\bf{A})\psi=0$ and $\bf{\hat{n}} \cdot \nabla \mu=0$. 
For the superconductor–metal interface, Neumann boundary conditions are applied to $\mu$ as $\partial\mu/\partial y = I$; while Dirichlet boundary conditions are imposed on $\psi$ as $\psi=0$,  preserving density continuity for both states~\cite{bishop2023pytdgl,jonsson2022current}.

\section{\label{app:Ic}Critical current}

\begin{figure}[!htbp]
    \centering
    \includegraphics[width=0.45\textwidth]{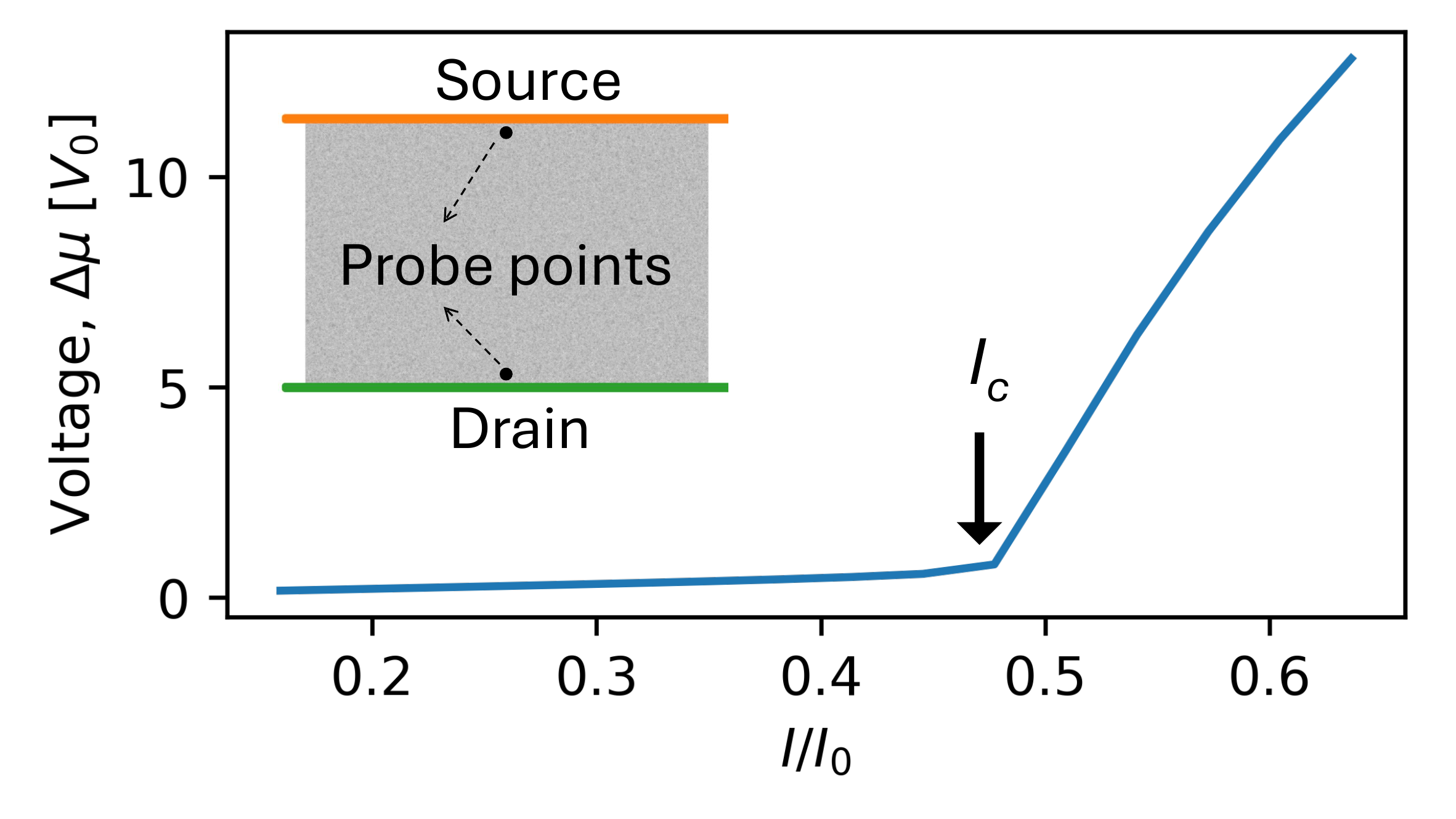}
    \caption{$I$–$V$ curve of the SC slab studied in this work. The inset illustrates the configuration of the voltage probes on the superconducting slab (gray area), including the source and drain terminals.}
    \label{fig:Ic}
\end{figure}

In superconductors, a driven current exceeding the critical current, and thus the local pair-breaking threshold, results in a strong suppression of the superconducting state. It can be observed and defined by the jump in the $I$-$V$ curve~\cite{berdiyorov2009kinematic}. In the gTDGL model, $I_c$ also serves as a criterion for studying current-driven vortex nucleation~\cite{berdiyorov2014dynamics}. Therefore, we adopt $I_c$ as the unit to reference the control parameters for vortex dynamics. Because the unit $I_0$ include the volume of sample,  the results not only dependent on parameters of material, but also determined by the overall geometry. In this simulation, the critical current $I_c$ is $0.48 I_0$, as demonstrated in ~\cref{fig:Ic}.

\section{\label{app:theta_s}Phase field of vortex pair}

This appendix provides the mathematical expression for the phase field profile $\theta_s$ of a VP, as illustrated in the insets of~\cref{fig:PF}(a). We consider an ideal vortex and an ideal antivortex located at positions $x_1 = \mathbf{r_V}$ and $x_2 = \mathbf{r_{aV}}$, respectively. Assuming each exhibits homogeneous rotational flow and follows a perfect circular profile of $\abs{\psi}$, the resulting phase field $\theta_s$ can be expressed as:
\begin{equation}
\label{eq:vortex}
\theta_{s,V}(\mathbf{r}) = +\tan^{-1}(\mathbf{r}-\mathbf{r_{V}})
\end{equation}
\begin{equation}
\label{eq:antivortex}
\theta_{s,aV}(\mathbf{r}) = -\tan^{-1}(\mathbf{r}-\mathbf{r_{aV}})
\end{equation}
The total phase field for a vortex–antivortex pair is then given by
\begin{equation}
\label{eq:VP}
\theta{s}(\mathbf{r}) = \theta_{V}(\mathbf{r}) + \theta_{aV}(\mathbf{r}) = \tan^{-1}(\mathbf{r} - \mathbf{r_V}) - \tan^{-1}(\mathbf{r} - \mathbf{r_{aV}}).
\end{equation}

For points far from both the vortex and antivortex ($|\mathbf{r}| \gg |\mathbf{r}_V|, |\mathbf{r}_{aV}|$), the two terms in ~\cref{eq:vortex} and ~\cref{eq:antivortex} nearly cancel, resulting in $\theta_{s}(\mathbf{r}) \approx 0$. Thus, the phase field is nearly constant outside the vortex–antivortex pair.

However, for points near VP, the phase field has changed dramatically. For instance, along the line perpendicular to the vortex and antivortex (i.e. $\mathbf{r} - \mathbf{r}_V = -(\mathbf{r} - \mathbf{r}_{aV})$), the phase field simplifies to
\begin{equation}
\theta_{s}(\mathbf{r}) = 2\tan^{-1}(\mathbf{r} - \mathbf{r}_V),
\end{equation}
which varies from $0$ to $2\pi$ as $\mathbf{r}$ moves across the pair, corresponding to a $2\pi$ phase jump across what we call the phase string. This result demonstrates that the characteristic $2\pi$ phase jump occurs between vortex and antivortex, while the phase remains nearly uniform outside.

\section{\label{app:VPEk} $E_k$ of vortex pairs}

The kinetic energy is given by
\begin{equation}
\label{eq:Ek_approx}
E_k = \int_{SC} da |(\nabla-i\mathbf{A})\psi|^2 \approx \abs{\psi}^2 \int_{SC} da |\nabla \theta_s|^2.
\end{equation}
Here we focus on the effect of the phase field and thus neglect density variations $\nabla\abs{\psi}$ arising from current-induced depairing of the superconducting state. In this work, the external DC current is applied along the $y$-direction, imposing a constant phase gradient described by $J_s \propto \partial_y \theta_s$. As a result, the $y$-component of the phase gradient, $\delta_y \theta_s$, dominates over the $x$-component, $\delta_x \theta_s$, in the term $\nabla \theta_s$. Therefore, in most of the discussions related to~\cref{fig:PF}, we focus on the analysis of $\delta_y \theta_s$. Notably, although a vortex–antivortex pair modifies $\theta_s$ only locally, it increases the density of phase contours along the $y$-direction, thereby extending its influence along the $y$-axis for $x$-oriented vortex pair.






\section{\label{app:Itr} Estimation of transition current}

In~\cref{fig:PF}(c), the threshold current $I_{tr}$ is estimated from the crossing point where $\delta_y \theta_s$ reaches $2\pi$. In this appendix, we provide additional details on how the $\delta_y \theta_s$ curves are obtained and analyzed.

\begin{figure}[!htbp]
    \centering
    \includegraphics[width=0.45\textwidth]{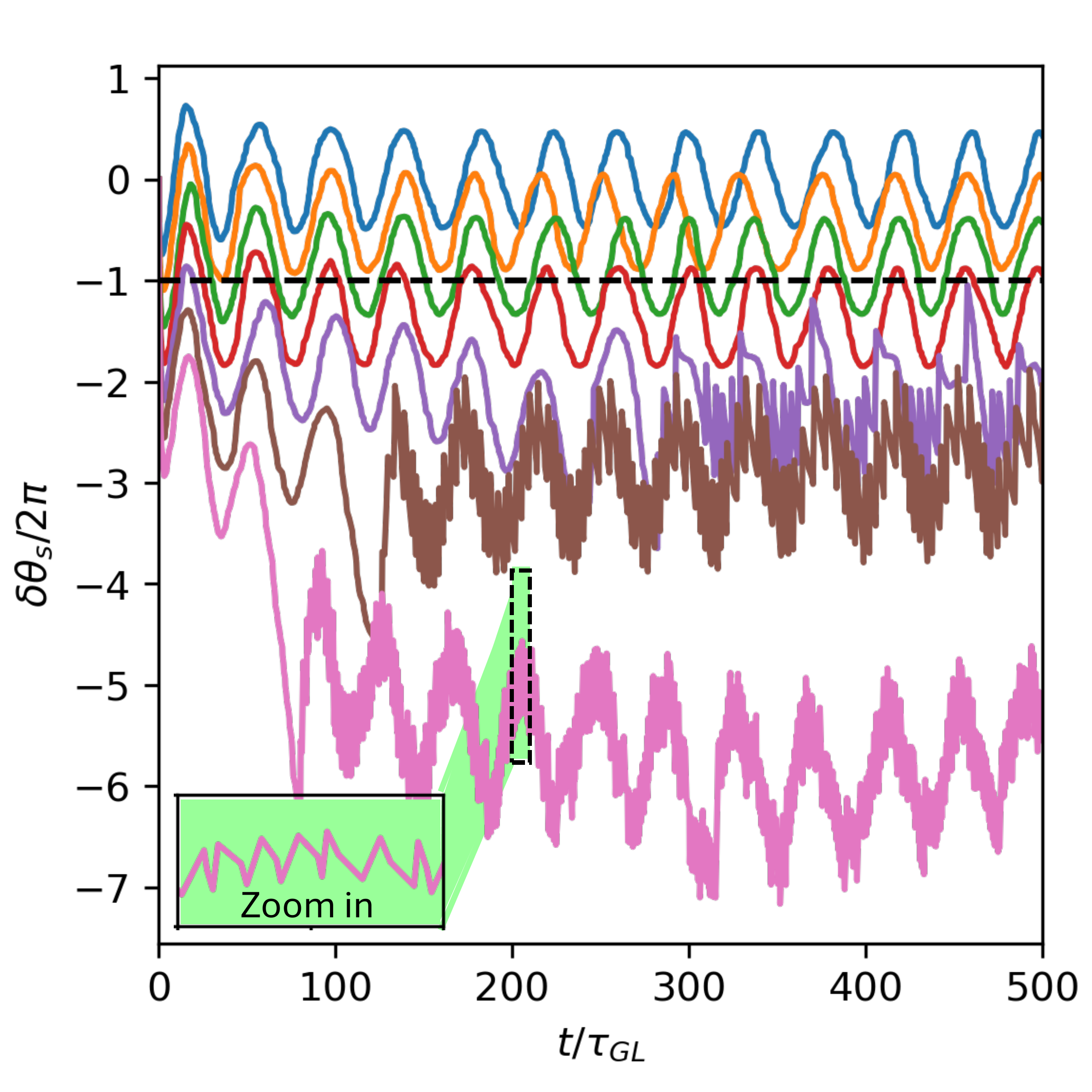}
    \caption{Time evolution of $\delta_y \theta_s$ discussed above. The current values corresponding to the time traces, from top to bottom, range from $0$ to $0.8I_c$. The inset shows a zoom-in view highlighting the vortex-crossing-induced zigzag oscillations in the range $t/\tau_{GL} \in [200, 210]$ and $\delta_y \theta_s/2\pi \in [-6, -4]$.  }
    \label{fig:Itr}
\end{figure}

For the curve of electric effect, the $\delta_y \theta_s$ is extracted from simulations with different $I$ and fixed $E = 0$. 
When including the optical effect in this estimation, the $\delta_y \theta_s$ induced by light cannot be neglected. As demonstrated in~\cref{fig:PF}(b), with non-zero $E$, $\delta_y \theta_s$  fluctuates in space and time due to light field. ~\cref{fig:Itr} shows time evolution of $\delta_y \theta_s$ traced far from the center ($x = 20\xi_0$) under varying current strengths, with fixed light intensity $E = 2E_0$, and the dashed line $\delta_y \theta_s = 2\pi$ denotes the topological restriction for vortex deconfinement. 
The bondary line indicated by the bracket that includes both electric and optical effects in~\cref{fig:PF}(b) is estimated from the minimum amplitude of the time trace shown in~\cref{fig:Itr}. The zigzag features observed in the curve (see zoom-in inset) arise from vortices crossing the tracking point during the deconfinement regime.

\section{\label{app:reflec_V} Boundary condition and vortex reflection}

The superconductor–vacuum interface ensures that neither supercurrent nor normal current flows across the boundary~\cite{jonsson2022current}. This physical requirement is implemented as a Neumann boundary condition for $\psi$, given by $\bf{\hat{n}} \cdot (\nabla-\it{i}\bf{A})\psi=(\nabla-\it{i}\bf{A})_x \psi=0$. Alternatively, this is equivalent to imposing a Dirichlet boundary condition on the supercurrent $J_s$:
\begin{equation}
J_{s,x} = \text{Im}[\psi^* (\nabla-\it{i}\bf{A})_x \psi]]=0,
\end{equation}
representing a fixed boundary for the supercurrent and a $180^\circ$ phase-shifted reflected wave.

At the boundary, the upper and lower sides of the vortex (with $\pm J_{s,x}$) change sign due to the Neumann condition, while the left and right sides retain their sign but are mirrored: the left side maps to the right, and vice versa, as a result of the reflection.  Consequently, the vorticity of the reflected vortex changes the sign under the Neumann boundary condition for $\psi$.

In most of cases with higher applied $I$  the phase slip occurs immediately after the vortex reaches the edge, subsequently inducing a $2\pi$ phase shift between the upper and lower regions of the sample (~\cref{fig:SS}(d)). This phenomenon has also been observed in studies of kinematic vortices in superconductors~\cite{berdiyorov2014dynamics,berdiyorov2009kinematic}. However, compared to conventional kinematic vortex-induced phase slips, the phase slip induced by optically assisted deconfinement of vortices requires a lower applied current. We are not in the regime where phase slips generate kinematic vortices. 
 
\end{appendices}

\clearpage
\bibliography{CDTSC}

\begin{thebibliography}{39}
\bibitem [{sup({\natexlab{b}})}]{supp_step}%
  \BibitemOpen
  \href@noop {} {} ({\natexlab{b}}),\ \bibinfo {note} {see {S}upplemental {M}aterial at URL-will-be-inserted-by-publisher for the video of simulation result.}\BibitemShut {Stop}%
\end{thebibliography}

\end{document}